\documentclass[reprint,10pt]{revtex4-1}

\usepackage{amsmath}    
\usepackage{graphicx}   
\usepackage{verbatim}   
\usepackage{color}      
\usepackage[hidelinks]{hyperref}   
\usepackage[caption=false]{subfig}
\begin{document}

\title{\bf{Extreme-ultraviolet-initiated High-harmonic Generation in Ar$^{+}$}}

\author{D. D. A. Clarke}
\email{dclarke23@qub.ac.uk}
\author{H. W. van der Hart}
\author{A. C. Brown}
\affiliation{Centre for Theoretical Atomic, Molecular and Optical Physics, School of Mathematics and Physics, \\
 Queen's University Belfast, University Road, Belfast, BT7 1NN, Northern Ireland}


\begin{abstract}
We employ the R-matrix with time-dependence method to investigate extreme-ultraviolet-initiated high-harmonic generation (XIHHG) in Ar$^{+}$.\ Using a combination of extreme-ultraviolet (XUV, $92\textrm{ nm}$, $3\times 10^{12}\,\textrm{Wcm}^{-2}$) and time-delayed, infrared (IR, $800\textrm{ nm}$, $3\times 10^{14}\,\textrm{Wcm}^{-2}$) laser pulses, we demonstrate that control over both the mechanism, and timing, of ionization can afford significant enhancements in the yield of plateau, and sub-threshold, harmonics alike.\ The presence of the XUV pulse is also shown to alter the relative contribution of different electron emission pathways.\ Manifestation of the Ar$^{+}$ electronic structure is found in the appearance of a pronounced Cooper minimum.\ Interferences amongst the outer-valence $3p$, and inner-valence $3s$, electrons are found to incur only a minor suppression of the harmonic intensities, at least for the present combination of XUV and IR laser light.\ Additionally, the dependence of the XIHHG efficiency on time delay is discussed, and rationalized with the aid of classical trajectory simulations.\
\end{abstract}

\maketitle

\section{Introduction}

High-harmonic generation (HHG) has emerged as the canonical representative for non-linear and ultrafast optical effects.\ The production of high-order harmonics of intense, femtosecond laser radiation, focused onto atomic or molecular gases, constitutes an indispensable capability for a myriad of scientific and technological endeavours.\ In particular, HHG has become a cornerstone of attosecond physics \cite{KrauszIvanov2009}, wherein its innately ultrafast nature has enabled the synthesis of coherent extreme-ultraviolet (XUV) and soft X-ray light pulses, with ever-shortening durations \cite{Goulielmakisetal2008,Zhaoetal2012,Popmintchevetal2012}.\

HHG is most simply understood in terms of the semiclassical three-step model \cite{Corkum1993}, a simple and physically appealing scheme that captures the gross features of the process.\ Here, an electron (i) tunnels through the laser-suppressed Coulomb barrier, (ii) is accelerated by the field, and (iii) radiatively recombines with the parent ion, all within a single cycle of the driving laser field.\

In more recent years, research in HHG has diversified beyond the development of light sources, acquiring new leitmotivs that reflect its potential as a sensitive probe of structure and dynamics.\ On the one hand, techniques of harmonic spectroscopy \cite{Smirnovaetal2009,Shineretal2011,Bruneretal2015} can facilitate the acquisition of intricate, field-free structural information.\ It is well-recognised, for instance, that HHG bears an intimate relation with the process of photoionization, and should thus encode details of the electronic structure of the irradiated target \cite{Woeneretal2009,Shineretal2012,Wongetal2013}.\ On the other hand, the development of advanced experimental techniques for assessing and controlling the spectral intensity \cite{Torresetal2007,Locketal2009}, phase \cite{Locketal2009,Mairesseetal2003}, polarization state \cite{Ferreetal2015,Lambertetal2015} and spatial divergence \cite{Farrelletal2011,Zairetal2013} of high-harmonic radiation has inspired novel approaches for monitoring photophysical processes in real-time.\ These include molecular structural changes \cite{Bakeretal2006} and charge migration \cite{Krausetal2015}, as well as chiral activity \cite{Ferreetal2015} on sub-femtosecond timescales.\

Concomitant with this growth of interest in the spectroscopic utilization of HHG has been a demand for enhanced control and optimization of the process, especially at the single-atom or single-molecule level.\ Tackling this demand necessitates a careful consideration of essentially all aspects of the three-step scheme, including the mechanism and timing of electron emission \cite{Gaardeetal2005}, the influence of multiple orbitals and ionization thresholds \cite{Smirnovaetal2009,BrownvanderHart2012}, the shape of the continuum electron wavepacket \cite{Frolovetal2009}, as well as the energy and orientation of the recollision event \cite{TelnovChu2009}.\ In particular, tunnel ionization from the initial state, traditionally recognised as the first step, yields only poor temporal control and a low conversion efficiency, being both confined to a fixed time interval during the laser pulse evolution, and restricted to the emission of an electron from the outermost atomic shell.\

Acknowledging these limitations, several authors have advocated schemes based on two-color laser fields \cite{Gaardeetal2005,Takahashietal2007,TudorovskayaLein2014}, whereby the target is subject to a strong visible or infrared (IR) driving pulse, as well as a short-duration, often somewhat weaker, XUV pulse.\ The HHG process can then be initiated by direct photoionization, or by excitation to a high-lying Rydberg state of the target, with subsequent tunnelling.\ Such alternatives to tunnel ionization, from the ground state, represent more efficient means of driving HHG in single atoms, offering improved control over the timing of the initial ionization event.\

The aforementioned schemes of XUV-initiated HHG (XIHHG) have constituted the focus of several experimental studies, exploring core excitation and correlated electron-hole dynamics in small molecules \cite{Leeuwenburghetal2013,Buth2015}.\ The three-step process has also been simulated in a number of theoretical works \cite{Biegertetal2006,Gademannetal2011}.\ More recently, XIHHG has been exploited to assess the relative contribution of inner- and outer-valence electrons to the harmonic response of Ne \cite{BrownvanderHart2016}.\ With a suitable combination of XUV and IR laser light, it was shown that the action of the 2{\it p} electrons may be selectively suppressed, and that of the 2{\it s} electrons thereby revealed.\ An important implication of this work is the possibility for HHG spectroscopy of more deeply bound electrons, and of the interference dynamics amongst inner- and outer-valence electrons.\

The potential of XIHHG, both as a more efficient mode of single-atom HHG, and as a spectroscopic tool, merits further investigation in general atomic and molecular systems.\ In this manuscript, we explore XIHHG in the Ar$^{+}$ ion, employing the {\it ab initio} R-matrix with time-dependence (RMT) methodology \cite{Mooreetal2011}.\ We devote particular attention to the role of an inner-valence excitation for this process.\ By tuning the XUV pulse appropriately, it is possible to excite the Ar$^{+}$ ion into a superposition of the $3s^{2}3p^{5} \, {^{2}P^{o}}$ ground and $3s3p^{6} \, {^{2}S^{e}}$ first excited states.\ Such tailoring of the initial state, with the potential to activate different ionization or recombination pathways, could represent a novel means of controlling the characteristics of the harmonic response.\

Effects arising from quantum interference, and their implications for atomic harmonic generation, have been treated in a number of theoretical works.\ It has been demonstrated, for instance, that if the system is prepared in a coherent superposition of different bound states \cite{Harris1989,MandelKoch1993,Gautheyetal1995}, in which ionization proceeds primarily from the most weakly bound, the harmonic spectrum can present two distinct plateaus, with different conversion efficiencies \cite{Watsonetal1996,Sanperaetal1996}.\ More recently, HHG from excited states was addressed experimentally, exploiting the attosecond lighthouse effect \cite{Beaulieuetal2016}.\ There, it was shown that population of high-lying Rydberg states can lead to either XUV free-induction decay \cite{Becketal2014,Campetal2015}, or near-threshold ionization followed by recombination to the ground state.\ We emphasize, however, that the present investigation differs from previous theoretical works in two key ways.\ Firstly, we consider the dynamics of a complex ion within a fully {\it ab initio} framework, as opposed to a few-level model \cite{Harris1989} or one-electron system \cite{Sanperaetal1996}.\ Secondly, we examine a plethora of additional effects that arise from XUV irradiation, otherwise absent in a single-color HHG scheme.\

More generally, atomic ions represent attractive targets for studies of HHG.\ They offer a stringent test for emerging theoretical techniques, whether in describing their field-free energy level structure, or field-driven response dynamics.\ Furthermore, higher electron binding energies, with respect to their neutral counterparts, imply increased cutoff energies, especially for the more strongly bound core electrons.\ Correspondingly, the harmonic spectra would present extended plateau structures, supporting the synthesis of ever-shorter attosecond pulses.\ The harmonic response of atomic ions has also proven of importance for interpreting experimental data:\ although experimental studies of HHG have addressed primarily neutral, noble gas targets, it has been suggested that the very highest-order harmonics detected arise from ionised species, generated during the intense laser-target interaction \cite{Gibsonetal2004}.\ Moreover, the prospect of harnessing HHG from ions, and thereby of extending frequency up-conversion techniques into still-shorter wavelength regimes, has been realised through the development of strategies for mitigating plasma-induced beam defocusing and phase mismatch \cite{Gibsonetal2004,Gaudiosietal2006} effects, which would otherwise limit the highest observable photon energies.\ It is thus of considerable practical interest to assess the HHG yields from noble gas ions, such as Ne$^{+}$ and Ar$^{+}$, both to inform, and direct, experimental HHG research.\

This manuscript is organized as follows.\ We begin with a brief summary of the RMT method.\ Following this, we examine the consequences of XUV multiphoton processes for IR-driven HHG in Ar$^{+}$.\ The possible role of the first excited state in providing an intermediate resonance, as well as of multielectron interference effects, involving the outer-valence $3p$ and inner-valence $3s$ electrons, are then discussed.\ To illustrate the importance of temporal control over the ionization event, we also address the dependence of the XIHHG efficiency on time delay, and supplement the analysis by means of classical trajectory simulations.\ The corresponding conclusions will close the manuscript.\

\section{Theory}

The study of ultrafast atomic dynamics in intense light fields demands sophisticated theoretical techniques, possessing the predictive capacity needed to explore multielectron correlations, and their consequences, in a first-principles manner.\ RMT theory offers an {\it ab initio} and non-perturbative technique for solving the time-dependent Schr\"{o}dinger equation (TDSE), appropriate to general multielectron atoms and atomic ions in strong laser fields.\ It represents the latest evolution in the development of a time-dependent R-matrix formalism \cite{BurkeBurke1997,Guanetal2007,Lysaghtetal2009}, whose flexibility and generality have been reflected in a plethora of recent applications.\ These include multielectron correlation in doubly and core-excited states of Ne \cite{Dingetal2016}, strong-field rescattering in F$^{-}$ \cite{Hassounehetal2015}, and spectral caustics in two-color HHG schemes \cite{Hamiltonetal2017}

A detailed exposition of RMT theory has been given in \cite{Mooreetal2011}, and so we merely provide a brief overview here.\ The method employs the traditional R-matrix paradigm of dividing configuration space into two separate regions.\ This partition is effected with respect to the radial coordinate of an ejected electron, and yields an inner region, containing the target nucleus, and an outer region of relatively large radial extent.\ Within the inner region, multielectron exchange and correlation effects are accounted for in the construction of the many-body wavefunction.\ In the outer region, the ionised electron is regarded as sufficiently distant from the residual ion that exchange may be neglected.\ This electron is thus subject only to the long-range, multipole potential of the residual system, as well as the applied laser field.\ Importantly, RMT relies on a hybrid numerical scheme, consisting of a unique integration of basis set and finite-difference techniques.\ This enables the most appropriate method for solving the TDSE to be applied in each region.\

In the inner region, the time-dependent, $N$-electron wavefunction is represented by an expansion in terms of R-matrix basis functions, where the time-dependence is incorporated in the expansion coefficients.\ These basis functions are generated from the $(N-1)$-electron wavefunctions of the residual ion states, as well as from a complete set of one-electron continuum functions describing the motion of the ejected electron.\ Additional $N$-electron correlation functions can be added to improve the quality of the basis set.\ The outer region wavefunction, in contrast, is constructed by means of the residual ion wavefunctions and radial wavefunctions of the ejected electron in each admissible channel.\ The latter are represented throughout a finite-difference grid, which not only affords a controllable discretization, but also a natural mode of parallelisation (through domain decomposition), which can be combined with other data decomposition approaches (appropriate for the inner region) in an efficient and scalable implementation.\ Continuity of the total wavefunction is enforced explicitly at the boundary, rather than via an R-matrix \cite{Mooreetal2011}.\ Indeed, the outer region grid is extended into the inner region, and the inner region wavefunction is evaluated at these interior points.\ This provides the boundary condition for the solution of the TDSE in the outer region.\ A derivative of the outer region wave function at the boundary is also made available to the inner region, enabling the inner region wave function to be updated.\

Whereas previous implementations of time-dependent R-matrix theory relied on the solution of a system of linear equations, and a low-order Crank-Nicolson propagator \cite{Lysaghtetal2009}, the RMT approach adopts a high-order Arnoldi scheme \cite{Smythetal1998}.\ This replaces the solution of a linear algebraic system with a series of matrix-vector multiplications, which may reduce the numerical error in both the temporal and spatial propagation of the wavefunction.\ Since the Arnoldi algorithm is dominated by such operations, the RMT methodology offers substantially improved parallel scalability, making feasible calculations that exploit massively parallel computing resources (with more than 1000 cores).\

Within the framework of RMT theory, harmonic spectra can be established following the method discussed in \cite{BrownRobinsonvdH2012}.\ Specifically, the spectra presented in this work are found from the expectation value of the dipole length operator,
\begin{equation}
{\bf d}(t) \propto \langle \Psi (t)\vert {\bf z} \vert \Psi (t) \rangle , \nonumber
\end{equation}
where ${\bf z}$ is the total position operator along the laser polarization axis, and $\Psi$ is the wavefunction.\ The harmonic yield is proportional to $\omega^{4}\vert {\bf d}(\omega)\vert^{2}$, where $\omega$ is the laser frequency and ${\bf d}(\omega)$ is the Fourier transform of ${\bf d}(t)$.\ However, as demonstrated in previous studies \cite{Hamiltonetal2017,BrownRobinsonvdH2012}, the spectra may also be ascertained through the expectation value of dipole velocity,
\begin{equation}
\dot{{\bf d}}(t) \propto \frac{d}{dt}\langle \Psi (t)\vert {\bf z} \vert \Psi (t) \rangle , \nonumber
\end{equation}
where the harmonic yield is proportional to $\omega^{2}\vert \dot{{\bf d}}(\omega)\vert^{2}$.\ The acceleration form, in contrast, is less appropriate for the RMT approach, as restrictions on the basis set mean that inner-shell electrons, such as the $1s$ electrons, are often maintained frozen.\ As a consequence, the calculations incorporate the action of the $1s$ electrons on those of the valence shell, but the back-action of the latter is not accounted for.\ This limitation precludes the use of the dipole acceleration in computing harmonic spectra for complex, multielectron atoms and ions.\ It should be noted that, independently of the manner in which the spectra are represented, we have consistently chosen to propagate the wavefunction in the length gauge.\ The latter has been found to give better results in conjunction with the description of atomic structure employed in time-dependent R-matrix calculations \cite{Hutchinsonetal2010}.\

\begin{figure}
\includegraphics[width=0.5\textwidth]{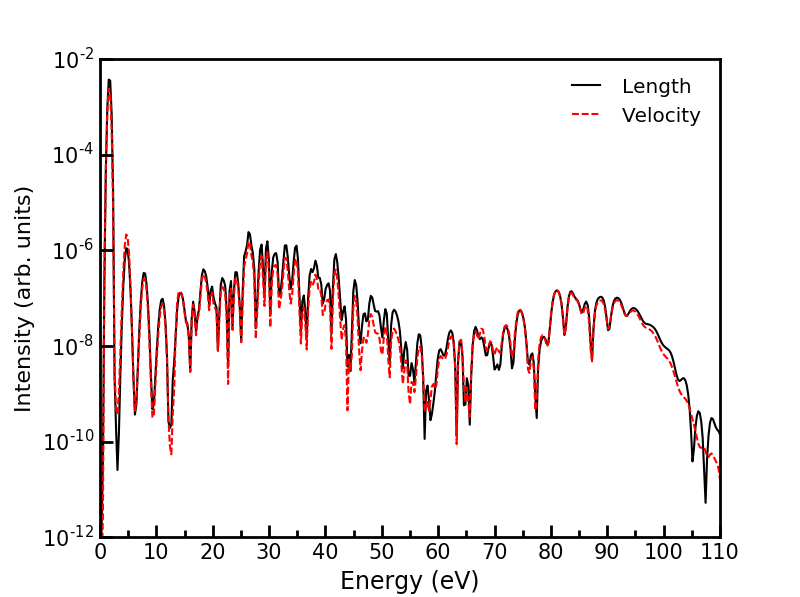}
\caption{(Color online) Harmonic spectra generated by the Ar$^{+}$ ion, subject to an $800\textrm{ nm}$, $3\times 10^{14}\,\textrm{Wcm}^{-2}$, IR laser pulse, as computed from the dipole length (solid, black line) and velocity (dashed, red line).}
\label{fig:two_gauges}
\end{figure}

The possibility of assessing the harmonic response, in either length or velocity form, offers an important check for accuracy within a simulation.\ As a relevant example, Figure \ref{fig:two_gauges} compares the harmonic spectra of Ar$^{+}$, subject to the IR field considered in this work, as computed from the dipole length and velocity.\ We observe only small discrepancies between the spectra, until well beyond the cutoff region (around $90 \textrm{ eV}$).\ Those differences, in turn, can be attributed to the limited basis set adopted for the Ar$^{2+}$ residual ion states (see Section \ref{sec:parameters}).\ Indeed, the present choice of basis ensures that the lowest ionization thresholds, corresponding to the Ar$^{2+}$ $3s^{2}3p^{4}\, {^{3}P^{e}}, {^{1}S^{e}}$ and ${^{1}D^{e}}$ states, are most faithfully represented.\ This facilitates good agreement between the spectra for energies up to approximately $30\textrm{ eV}$, where their associated Rydberg series are active.\ More significant deviations are found in the range $35\textrm{ eV}$ to $55\textrm{ eV}$, which is dominated by excitations from the $3s$ subshell.\ Further expansion of the basis set, to include additional, suitably optimized orbitals, might improve the description of the Ar$^{2+}$ $3s3p^{5}\, {^{3}P^{o}}$ and ${^{1}P^{o}}$ states, and provide still better agreement among the spectra in this region.\ Notwithstanding these differences, the generally high degree of consistency, among the spectra, is suggestive of the numerical quality of the wavefunction data employed in the present calculations.\

To supplement analysis of the harmonic spectra, we perform classical trajectory simulations based on the three-step model \cite{Corkum1993}.\ We assume that an electron is tunnel-ionized into the continuum with zero initial velocity, and for each possible ionization time, the electron velocity and position are determined through numerical solution of the classical (Newtonian) equation of motion.\ Those electrons which recollide, and thereby elicit high-harmonic emission, describe trajectories which pass through the origin.\ The energy of the harmonic photons is then readily calculated from the electron recollision energy and the ionization potential.\ Whilst such a model possesses obvious limitations, we find it to be a particularly simple and convenient means of estimating harmonic cutoff energies, for both single and multi-color field configurations.\

\section{Calculation Parameters}
\label{sec:parameters}

The Ar$^{+}$ target considered in this work is as discussed in previous R-matrix studies \cite{BrownvanderHart2012,BrownvanderHart2013}.\ Within the inner region, we regard the ion as Ar$^{2+}$ to which is added a single electron.\ To describe the structure of Ar$^{2+}$, we employ a set of Hartree-Fock $1s,2s,2p,3s,3p$ orbitals, acquired for the Ar$^{2+}$ ground state from the data of Clementi and Roetti \cite{ClementiRoetti}.\ The $3s^{2}3p^{4}\, {^{3}P^{e}}, {^{1}S^{e}}, {^{1}D^{e}}$ and $3s3p^{5}\, {^{3}P^{o}}, {^{1}P^{o}}$ states of Ar$^{2+}$ are obtained in the form of configuration-interaction expansions, comprising the $3s^{2}3p^{4}$, $3s3p^{5}$ and $3p^{6}$ configurations.\ Such an approach permits flexibility in the degree of atomic structure included in the calculations, which has proven especially important in assessing the relative contribution of different electron emission channels to the harmonic spectrum \cite{BrownvanderHart2012,BrownvanderHart2013}.\ The initial state is the Ar$^{+}$ $3s^{2}3p^{5} \, {^{2}P^{o}}$ ground state, with total magnetic quantum number $M=0$.\ This corresponds to the dominant (non-relativistic) Ar$^{+}$  ground state level following strong-field ionization of Ar at short wavelengths.\

The radial extent of the inner region is $20\textrm{ a.u.}$, which suffices to effectively confine the orbitals of the residual Ar$^{2+}$ ion.\ The inner region continuum functions are generated using a set of 60 B-splines of order 13 for each available orbital angular momentum of the outgoing electron.\ We retain all admissible $3s^{2}3p^{4}\epsilon l$ and $3s3p^{5}\epsilon l$ channels up to a maximum total orbital angular momentum $L_{\textrm{max}} = 70$.\ The outer region boundary radius is $2000\textrm{ a.u.}$, ensuring that no unphysical interference structure in the wavefunctions arise through reflection of the ejected electron wavepacket.\ The finite-difference grid spacing is $0.08\textrm{ a.u.}$, and a time-step of $0.01\textrm{ a.u.}$ is adopted for propagation of the wavefunction.\

We employ a two-color irradiation scheme, comprising a four-cycle XUV pulse, with a peak intensity of $3\times 10^{12}\,\textrm{Wcm}^{-2}$ and a central wavelength of $92.0\textrm{ nm}$  (or a central photon energy of $\hbar\omega_{\textrm{XUV}}\approx 13.5\textrm{ eV}$), and a time-delayed, six-cycle IR pulse, with peak intensity $3\times 10^{14}\,\textrm{Wcm}^{-2}$ and central wavelength $800\textrm{ nm}$.\ The former is responsible for initiating an excitation and/or ionization response from the Ar$^{+}$ target.\ In particular, a one-photon absorption event can excite the ion into a superposition of the $3s^{2}3p^{5} \, {^{2}P^{o}}$ ground and $3s3p^{6} \, {^{2}S^{e}}$ first excited states.\ The latter, in contrast, mediates the electron recollision dynamics, thereby eliciting high-harmonic emission.\ Both pulses are linearly polarized along the $z$-axis, and are assumed to exhibit a sine-squared ramp-on/off temporal profile.\ The time delay, $\Delta$, is measured between the central peaks of the two pulses, and is always chosen such that the XUV pulse attains peak intensity before the IR pulse.\

\section{Results}

\subsection{Two-color Harmonic Spectrum}

To assess the features of XIHHG from Ar$^{+}$, we consider the harmonic spectra produced upon subjecting the ion to both single-color (IR only) and two-color (IR + XUV) field configurations.\ For the latter, we assume a fixed time delay between the pulses, such that the XUV peak is near-coincident in time with the penultimate IR peak.\ This corresponds to a delay of approximately $\Delta = 3.5T_{\textrm{XUV}}$, where $T_{\textrm{XUV}} = 2\pi/\omega_{\textrm{XUV}}$ is the XUV pulse period and $\omega_{\textrm{XUV}}$ the central frequency.\ According to the three-step model, electrons released into the field at this time describe the optimal trajectories for HHG.\ We thus expect any promotion of ionization, afforded by the XUV pulse, to be manifested most clearly under these conditions.\

\begin{figure}
\includegraphics[width=0.5\textwidth]{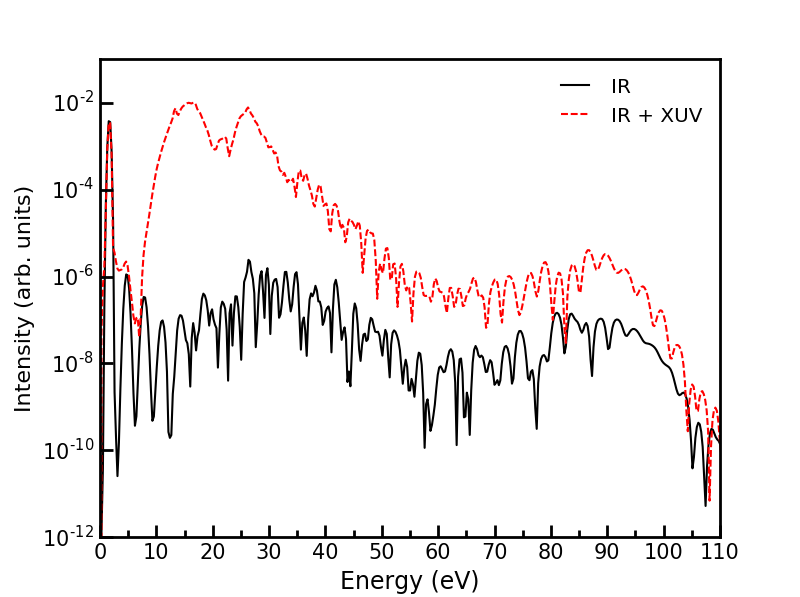}
\caption{(Color online) Harmonic spectra generated by the Ar$^{+}$ ion, subject to single-color (solid, black line) and two-color (dashed, red line) laser field configurations.\ The single-color field comprises an $800\textrm{ nm}$, $3\times 10^{14}\,\textrm{Wcm}^{-2}$, IR laser pulse.\ The two-color scheme consists of the same IR pulse, in combination with a $92\textrm{ nm}$, $3\times 10^{12}\,\textrm{Wcm}^{-2}$, XUV pulse.\ The latter is timed to coincide with the penultimate maximum of the IR field.}
\label{fig:two_color}
\end{figure}

Figure \ref{fig:two_color} displays the single- and two-color harmonic spectra.\ Perhaps the most striking consequence of XUV irradiation is an enhancement, by up to four orders of magnitude, in the yield of low-energy harmonics over the single-field case.\ This effect is particularly noteworthy at energies comparable to both $\hbar\omega_{\textrm{XUV}}$ (unsurprisingly), and to the ionization threshold for the $3p$ subshell $(I_{3p}\approx 27.6\textrm{ eV})$.\ A sustained enhancement, by as much as two orders of magnitude, is also found throughout the plateau region, which spans around $60\textrm{ eV}$.\

The high-order harmonics are of particular interest, given their role in frequency up-conversion applications.\ We attribute the increased response above the ionization threshold to multiphoton processes, mediated by one or both laser fields.\ Indeed, absorption of two XUV photons can excite the ion into a Rydberg state of the series converging onto any of the $3p$ thresholds.\ This dramatically increases the susceptibility of the ion to tunnel ionization in the IR field.\ Such a two-stage process enables the electron to be released into the field with approximately zero energy, which is desirable for optimizing the three-step recollision process.\ The finite bandwidth of the XUV pulse (approximately $6.8\textrm{ eV}$), however, may also support direct photoionization from the $3p$ subshell, through absorption of at least two photons.\ This is also expected to contribute to the increased response at threshold.\ Note that the Ar$^{+}$ $3s3p^{6} \, {^{2}S^{e}}$ state, accessible by one-photon absorption from the XUV field, may provide a resonant enhancement of these multiphoton processes, whether they result in Rydberg state excitation, or direct ionization to the continuum.\ However, we must establish confirmation that this state truly participates in the dynamics, before making such an assertion (see Section \ref{subsec:excited}).\

The aforementioned XUV-dependent mechanisms of ionization also influence the harmonic yield in the cutoff region.\ For the single-field spectrum, we find the harmonic cutoff to be consistent with the semiclassically predicted value \cite{Krauseetal1992},
\begin{equation}
E_{\textrm{co}}^{\textrm{IR}} \approx 1.2I_{3p} + 3.2U_{\textrm{p}} \approx 90.5\textrm{ eV}, \nonumber
\end{equation}
where $U_{\textrm{p}}$ is the ponderomotive energy, and we have assumed ionization from the $3p$ subshell.\ The comparable cutoff energy, observed for the two-color spectrum, suggests that the highest-energy harmonics there are also determined by the action of a $3p$ electron, with the recollision dynamics driven almost solely by the IR field.\ The choice of time delay ($3.5T_{\textrm{XUV}}$, or approximately one half-cycle of the IR pulse) ensures that ionization, or increased susceptibility to it, is realised within a temporal interval in which the IR pulse attains its penultimate peak.\ As emphasized previously, an electron ejected within this interval will not only recollide with the parent ion, but do so with the maximum possible kinetic energy acquired under the given field conditions.\ We can therefore rationalize the higher yield of cutoff harmonics in terms of enhanced ionization from the $3p$ subshell, driven directly by or assisted with the XUV pulse, within an optimal sub-cycle time period.\

The choice of XUV photon energy ($\hbar\omega_{\textrm{XUV}}\approx 13.5\textrm{ eV}$) was made with the aim of probing the $3s3p^{6} \, {^{2}S^{e}}$ state, and we discuss the importance of this excitation in Section \ref{subsec:excited}.\ The present XIHHG scheme thus requires the absorption of at least two XUV photons to excite a $3p$ electron close to the ionization threshold ($2\hbar\omega_{\textrm{XUV}}\approx I_{3p}$).\ It is natural to question, however, if the enhanced harmonic response, observed in Figure \ref{fig:two_color}, could have been achieved through a single-photon process alone.\ Indeed, we could envisage direct excitation of Ar$^{+}$ from the ground state, to either a weakly bound Rydberg state, or even to the continuum, by the absorption of a single, sufficiently energetic photon $(\hbar\omega_{\textrm{XUV}}\approx I_{3p})$, with otherwise little selectivity warranted in the choice of pulse frequency.\ Such a scheme would be somewhat akin to that proposed by Brown and van der Hart \cite{BrownvanderHart2016}, who employed below-threshold, XUV laser light to populate selected Rydberg series of Ne.\

To address this question, we have ascertained the two-color spectrum in the case $\hbar\omega_{\textrm{XUV}}\approx 27.0\textrm{ eV}$ (not shown).\ The results suggest that, whilst the qualitative features of the spectrum remain unchanged above the $3p$ ionization threshold, the yield of high-order harmonics can be raised by as much as an order of magnitude.\ The latter can be understood most simply on the basis of relative probabilities, pertaining to XUV single- and multiphoton processes:\ a one-photon absorption event is more probable than any higher-order process, in which two or more photons are absorbed.\ A further advantage, offered by this one-photon XIHHG strategy, is that enhanced harmonic yields could likely be realised with still-lower XUV peak intensities.\

\subsection{Competition amongst the $3s^{2}3p^{4}\, {^{3}P^{e}}, {^{1}S^{e}}$ and ${^{1}D^{e}}$ Thresholds}
\label{subsec:competition}

Previous R-matrix studies \cite{BrownvanderHart2012,BrownvanderHart2013} have shown that, for single-field harmonic generation in Ar$^{+}$, aligned with $M=0$, channels associated with the ${^{1}S^{e}}$, and especially ${^{1}D^{e}}$, thresholds are most influential for the harmonic yield.\ Tunnel ionization, leaving Ar$^{2+}$ in the ground ${^{3}P^{e}}$ state, was found to contribute much less significantly to the total yield.\ We emphasize that the distribution of single-electron magnetic quantum number values, $m$, plays a fundamental role in suppressing the ${^{3}P^{e}}$ channels in an IR-only HHG scheme.\ The ${^{3}P^{e}}$ threshold is unavailable for emission/excitation of an electron with $m=0$, requiring the transition of one with $m = \pm 1$.\ Such a condition will be most restrictive during the initial ionization phase of the three-step mechanism, where IR photon absorption will favour excitation from an orbital aligned along the laser polarization axis $(m=0)$.\ As such, electrons with $m = \pm 1$ exhibit a weak ionization response, and channels associated with the ${^{3}P^{e}}$ threshold are poorly populated.\

The aforementioned works treated exclusively the dynamics of HHG in a single-color field.\ In contrast, we have here chosen to enhance the conversion efficiency by means of an additional XUV pulse, which provides an initial excitation of the electron to be ejected.\ The question arises as to whether XUV irradiation might also affect the competition amongst low-lying ionization thresholds, especially of the ${^{3}P^{e}}$ and ${^{1}D^{e}}$ thresholds in the Ar$^{+}$ harmonic response.\ To reveal their interplay in the present XIHHG scheme, we have effected a change in the field-free, Ar$^{+}$ electronic structure, selectively removing each of the two thresholds, and computing the harmonic spectrum in each case.\

\begin{figure}
\includegraphics[width=0.5\textwidth]{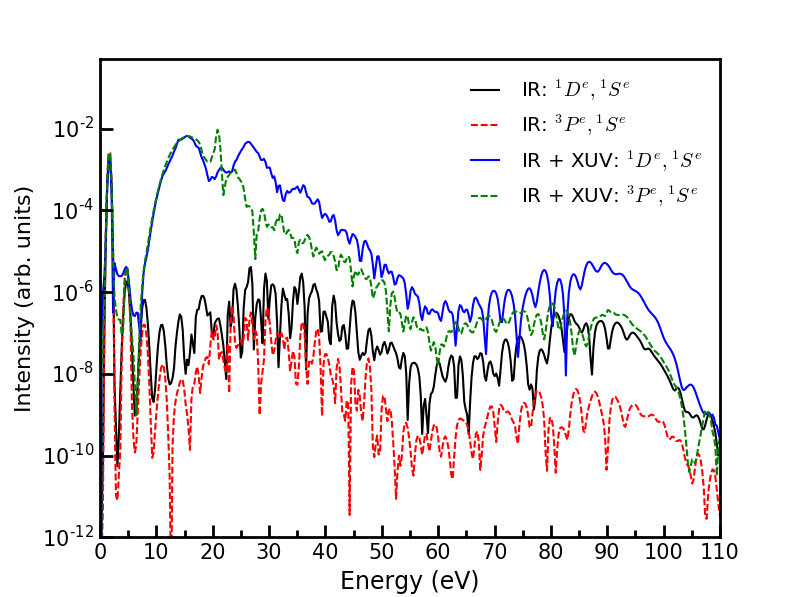}
\caption{(Color online) Harmonic spectra generated by the Ar$^{+}$ ion, subject to single-color and two-color laser field configurations.\ The single-color field comprises an $800\textrm{ nm}$, $3\times 10^{14}\,\textrm{Wcm}^{-2}$, IR laser pulse.\ The two-color scheme consists of the same IR pulse, in combination with a $92\textrm{ nm}$, $3\times 10^{12}\,\textrm{Wcm}^{-2}$, XUV pulse.\ The latter is timed to coincide with the penultimate maximum of the IR field.\ The yields are compared for different descriptions of the Ar$^{+}$ structure, incorporating the Ar$^{2+}$ $3s^{2}3p^{4}$ ${^{1}S^{e}}$ and ${^{1}D^{e}}$ thresholds (lower, solid black line and upper, solid blue line), or ${^{3}P^{e}}$ and ${^{1}S^{e}}$ thresholds (lower, dashed red line and upper, dashed green line).}
\label{fig:thresh_low}
\end{figure}

Figure \ref{fig:thresh_low} evidences the effect of restricting the number of ionization thresholds on the HHG yield, for both single- and two-color field configurations.\ For clarity, we have omitted the full spectrum pertaining to each of these (but shown in Figure \ref{fig:two_color}), which account for the influence of all three $3p$ ionization thresholds $({^{3}P^{e}}, {^{1}S^{e}}, {^{1}D^{e}})$.\ Indeed, we have found the latter spectra to be well-approximated through inclusion of only the Ar$^{2+}$ ${^{1}S^{e}}$ and ${^{1}D^{e}}$ thresholds, suggesting that, as in single-field HHG \cite{BrownvanderHart2012,BrownvanderHart2013}, channels connected to these excited residual ion states are dominant.\ Comparison of the two-color $({^{3}P^{e}}, {^{1}S^{e}})$ and $({^{1}D^{e}}, {^{1}S^{e}})$ spectra provides additional insight into the role of the ${^{1}D^{e}}$ threshold.\ We find that its removal reduces the yield of plateau harmonics by up to an order of magnitude, principally mitigating the ehancements (observed in Figure \ref{fig:two_color}) in the cutoff region, as well as at energies comparable to $I_{3p}$.\ Interestingly, however, the presence of the XUV field actually improves the relative contribution of the ${^{3}P^{e}}$ threshold, by more than a factor of 5 for harmonic photon energies exceeding $40\textrm{ eV}$.\

We claim that the presence of an additional XUV pulse alters the relative importance of $m =0$ and $m = \pm 1$ electron emission pathways.\ An XUV field interacts with both $m =0$ and $m = \pm 1$ electrons of the valence $3p$ subshell, albeit more strongly with the former.\ Absorption of one or more XUV photons can promote an electron to some higher energy state, giving rise to a Rydberg state of the ion, or some virtual state.\ Once an excitation has been realised, the electron can interact more freely with the IR field, so that the initial $m$ value (orbital alignment) plays a reduced role.\ We therefore expect that the emission pathway for $m = \pm 1$ electrons will be enhanced when ionization proceeds from an XUV-excited state, rather than from the ground state (as in single-color HHG).\ This would reduce the discrepancy between the contributions of $m=0$ and $m=\pm 1$ electron emission for a two-color scheme, reflected in an increased yield arising from the ${^{3}P^{e}}$ channels.\


\subsection{Cooper Minimum}

The single-field spectrum of Figure \ref{fig:two_color} displays a rather prominent minimum, localized to the energy range $40\textrm{ eV}$ to $80\textrm{ eV}$.\ This feature persists in the two-color spectrum also.\ In general, the presence of minima in atomic or molecular harmonic spectra can be attributed to any one of several possible effects.\ Perhaps the most actively studied are those that pertain to the intrinsic structure of the irradiated target, whether electronic (such as the well-known Cooper minimum \cite{Woeneretal2009,Higuetetal2011}) or geometric (as in molecular multi-centre interference \cite{Manschwetusetal2015}).\ Recent work has suggested, however, that suppression of the harmonic yield can also result from the field-driven, post-ionization dynamics of the ejected electron or residual system \cite{Kanaietal2007}.\ A crucial discriminant between structure- and dynamically-derived effects, as already discussed by W\"{o}rner {\it et al.}\ \cite{Woeneretal2009}, is that the latter should present a sensitive dependence on irradiation conditions, with their characteristic spectral features being controlled, in particular, by the laser light intensity and wavelength.\

\begin{figure}
\includegraphics[width=0.5\textwidth]{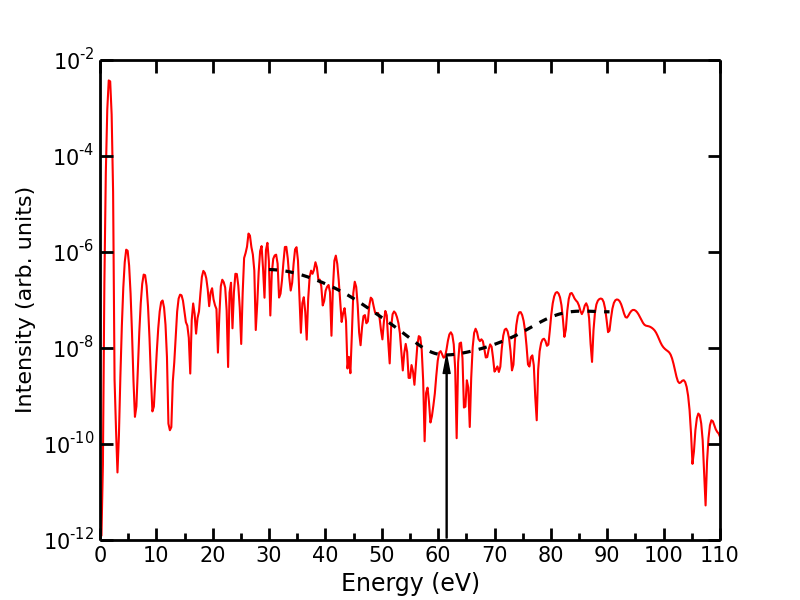}
\caption{(Color online) Harmonic spectrum generated by the Ar$^{+}$ ion, subject to an $800\textrm{ nm}$, $3\times 10^{14}\,\textrm{Wcm}^{-2}$, IR laser pulse (solid, red line), with Gaussian smoothing (dashed, black line).\ The arrow indicates the position of the minimum, at approximately $61.5\textrm{ eV}$.}
\label{fig:smooth}
\end{figure}

To constrain the possible origins of the minimum in the single-field spectrum of Ar$^{+}$, we have performed a series of calculations for a range of IR pulse peak intensities and wavelengths.\ In each case, and following Higuet {\it et al.}\ \cite{Higuetetal2011}, we estimate the position of the minimum by means of a Gaussian filter, as demonstrated in Figure \ref{fig:smooth} for the IR-only spectrum.\ There, we locate the feature at $61.5\textrm{ eV}$.\

\begin{figure}[h]
\includegraphics[width=\columnwidth]{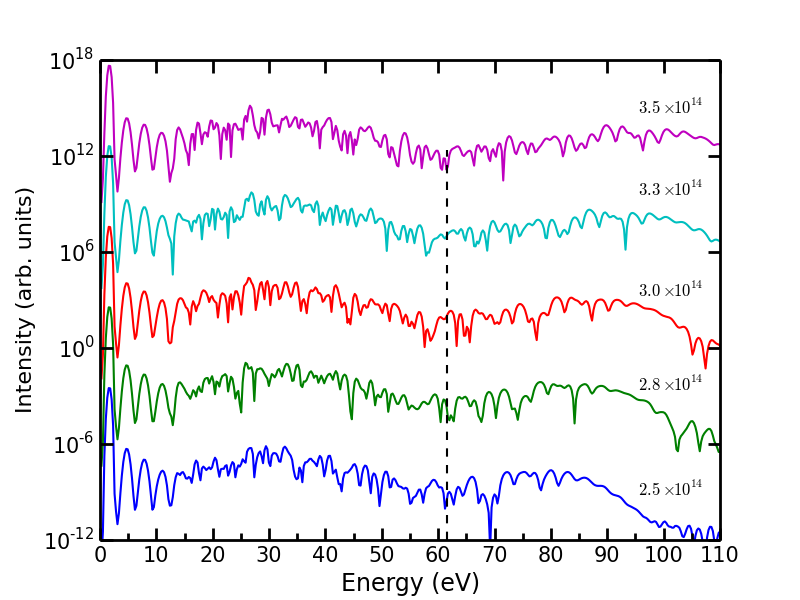}
\caption{(Color online) Harmonic spectra generated by the Ar$^{+}$ ion, subject to an $800\textrm{ nm}$, IR laser pulse of variable peak intensity.\ The spectra are successively offset by a factor of $10^{5}$, and all intensities are expressed in units of $\textrm{Wcm}^{-2}$.\ The approximate position of the minimum, which varies weakly about $61.5\textrm{ eV}$, is indicated by the dashed line.}
\label{fig:position_intensity}
\end{figure}

\begin{figure}[h]
\includegraphics[width=\columnwidth]{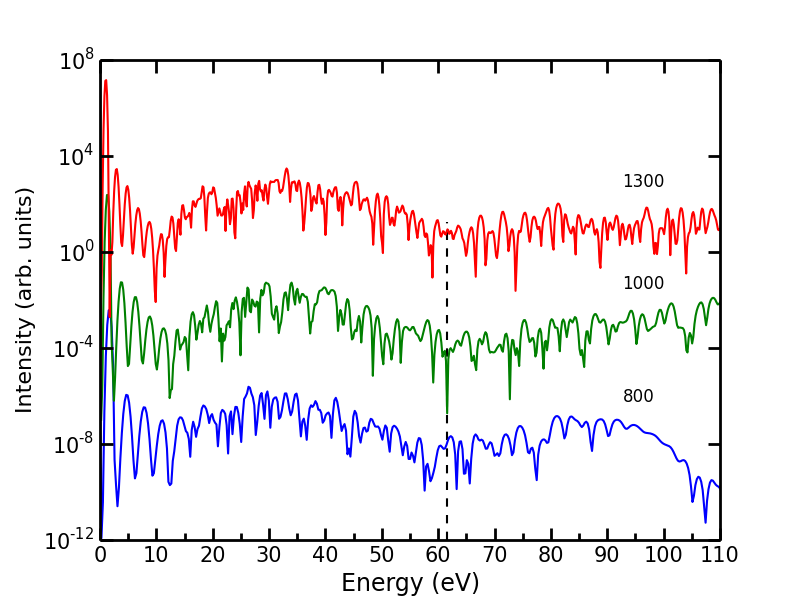}
\caption{(Color online) Harmonic spectra generated by the Ar$^{+}$ ion, subject to a $3\times 10^{14}\,\textrm{Wcm}^{-2}$, IR laser pulse of variable wavelength.\ The spectra are successively offset by a factor of $10^{5}$, and all wavelengths are expressed in units of nm.\ The approximate position of the minimum, which varies weakly about $61.5\textrm{ eV}$, is indicated by the dashed line.}
\label{fig:position_wavelength}
\end{figure}

The influence of laser intensity on the Ar$^{+}$ spectrum is demonstrated in Figure \ref{fig:position_intensity}.\ As expected, an increase in peak intensity effects both a rise in the cutoff energy, as well as in the global harmonic yield.\ Crucially, the position of the minimum remains largely unaffected, with shifts that are both unsystematic, and typically of a sub-electron-volt scale.\ Similarly, in Figure \ref{fig:position_wavelength}, we consider possible changes in the location of the minimum for three distinct IR wavelengths.\ Note that those shorter than $800\textrm{ nm}$, for the fixed intensity of interest $(3\times 10^{14}\,\textrm{Wcm}^{-2})$, reduce the harmonic cutoff energy sufficiently that the minimum no longer appears well-defined (if at all).\ We thus restrict the comparison of harmonic yields to those at longer wavelengths.\ Despite an increase of over $50\%$ in this quantity, and correspondingly of almost $69\textrm{ eV}$ in the ponderomotive energy, we find comparatively minor shifts in the minimum.\ The general robustness of this feature, with respect to changes in irradiation conditions, favours the conclusion that it must be associated, in some fashion, with the electronic structure of the Ar$^{+}$ ion, as opposed to the dynamics of laser-driven recollision.\

Perhaps the most plausible explanation for this feature, given the Ar-like nature of the system, is that the suppression of yield reflects the emergence of a Cooper minimum \cite{Woeneretal2009,Higuetetal2011}.\ This is supported by recent experimental \cite{Covingtonetal2011} and theoretical \cite{Tyndalletal2016} investigations of the photoionization spectrum of Ar$^{+}$, which clearly demonstrate the existence of such a feature.\ Importantly, however, the minimum observed in photoionization spectra arises at a somewhat lower energy, in the range $45\textrm{ eV}$ to $50\textrm{ eV}$.\ This is in line with studies of the Cooper minimum in neutral Ar, which has been reported to systematically shift between harmonic and photoionization spectra \cite{Higuetetal2011}.\ Due to the paucity of experimental data for HHG in Ar$^{+}$, it is difficult to state conclusively that such a shift also occurs for this species.\ More significantly, theoretical predictions for such electronic structure features can be rather sensitive to the quality of atomic wavefunctions.\ In particular, a faithful representation of low-lying residual ion states is essential for the accuracy of photoionization data.\ Previous time-independent R-matrix studies \cite{Tyndalletal2016} have suggested that elaborate Ar$^{2+}$ descriptions, employing large numbers of configurations and relativistic corrections to the Hamiltonian, are necessary for a reliable account of Ar$^{+}$ single-photon photoionization, even from the valence shell.\ Thus, whilst the limited degree of atomic structure, adopted in this work, is sufficient to identify the most important facets of the XIHHG process, quantitative predictions for such features, depending very sensitively on that structure, should be made with care.

\subsection{Role of the $3s3p^{6} \, {^{2}S^{e}}$ Excited State}
\label{subsec:excited}

As mentioned previously, the choice of XUV photon energy ($\hbar\omega_{\textrm{XUV}}\approx 13.5\textrm{ eV}$) was made with the aim of probing the $3s3p^{6} \, {^{2}S^{e}}$ excited state.\ The two-color spectrum of Figure \ref{fig:two_color} presents a broad peak structure at energies comparable to the $3s^{2}3p^{5}\, {^{2}P^{o}}\rightarrow3s3p^{6} \, {^{2}S^{e}}$ transition energy.\ To establish whether the excited ${^{2}S^{e}}$ state truly plays a mediating role in the present two-color HHG process, we must examine its contribution in greater detail.\

Supplementary RMT calculations (results not shown) reveal that this state is only weakly populated during XUV irradiation, with the maximum population attained of order $10^{-3}$.\ Systematic detuning from the expected resonance, through variation of the XUV central photon energy by $\pm 10\textrm{ eV}$, yields only a simple reduction in the occupancy for both positive and negative changes in $\hbar\omega_{\textrm{XUV}}$.\ We note that this behaviour points to the absence of a strong ponderomotive shift in the ${^{2}S^{e}}$ state, which would otherwise ensure its resonant population at a somewhat higher or lower XUV frequency.\ Moreover, by selectively removing this state from the dynamics, we have found only insignificant alterations in the aforementioned peak structure of Figure \ref{fig:two_color}.\ We therefore conclude that excitation of the Ar$^{+}$ $3s3p^{6}\, {^{2}S^{e}}$ state is inconsequential for the present XIHHG scheme.\

A plausible explanation, for the apparent insignificance of the ${^{2}S^{e}}$ intermediate resonance, may lie in the relative magnitudes of the dipole matrix elements for transitions amongst different low-lying, bound states of Ar$^{+}$.\ Indeed, time-independent R-matrix calculations for the static Ar$^{+}$ structure, performed by the present authors, indicate that the dipole matrix element for a $3s\rightarrow 3p$ transition can be up to an order of magnitude lower than those relevant for other transitions, including $3p\rightarrow 4s$, $3p\rightarrow 3d$, $3d\rightarrow 4p$ and $4s\rightarrow 4p$.\ These predictions are supported by the data of both Hibbert and Hansen \cite{HibbertHansen1994} and Berrington {\it et al.}\ \cite{Berringtonetal2001}, which indicate that the corresponding discrepancy in oscillator strengths can readily exceed an order of magnitude.\ This would suggest that the first excited $({^{2}S^{e}})$ state of Ar$^{+}$ may not be optimal for realising resonantly enhanced, multiphoton ionization yields in XIHHG schemes, or of driving HHG from an excited state.\ In connection with the latter, alternative strategies based on XUV free-induction decay \cite{Becketal2014,Campetal2015,Beaulieuetal2016}, or direct ionization from an excited state followed by recombination to the ground state \cite{Beaulieuetal2016}, offer considerable promise.

\subsection{$3s$ versus $3p$ Emission}

A long-standing feature of time-dependent R-matrix techniques has been their capacity for exploring the correlated response of multiple electrons in laser-driven atomic processes.\ For instance, the impact of window resonances on low-order harmonic generation from Ar has been described in a first-principles manner \cite{Brownetal2012}.\ More recently, the predictive capabilities of the RMT methodology were exploited in a study of XIHHG from neon \cite{BrownvanderHart2016}.\ There, a novel scheme had been proposed for resolving the contribution of inner- and outer-valence electrons to the harmonic response, with implications for HHG spectroscopy of their interference dynamics.\

Given the rich behaviour observed in these studies, and notwithstanding the conclusions of the preceding section, it is natural to question if any signatures of a $3s$-electron response are manifest in the present XIHHG scheme.\ To this end, we find that even a gross analysis of the two-color spectrum (Figure \ref{fig:two_color}) yields some insight.\ The pronounced intensity of harmonics, around the $3p$ ionization threshold, suggests that the outer-valence electrons dominate the ionization yield.\ The higher binding energy of $3s$ electrons would also give rise to a double-plateau structure, with a second cutoff feature (above $110\textrm{ eV}$).\ Since the two-color spectrum displays only a single cutoff, in the energy range expected for $3p$ electron dynamics alone, we cannot claim that the $3s$ electrons elicit the generation of the highest observable harmonics.\

In order to isolate the contribution of the $3s$ electrons more effectively, and thereby conduct a more systematic assessment of their role, we modify the present description of the field-free, Ar$^{+}$ electronic structure.\ More specifically, we selectively remove the $3s3p^{5}\, {^{3}P^{o}}$ and $3s3p^{5}\, {^{1}P^{o}}$ Ar$^{2+}$ ionization thresholds, tantamount to neglecting ionization from the $3s$ subshell.\ Note that a similar approach had been adopted to investigate the contribution of the $3s$ and $3p$ electrons to low-energy, single-field harmonic generation in both Ne$^{+}$ \cite{Hassounehetal2014} and Ar$^{+}$ \cite{BrownvanderHart2012}.\

\begin{figure}
\includegraphics[width=\columnwidth]{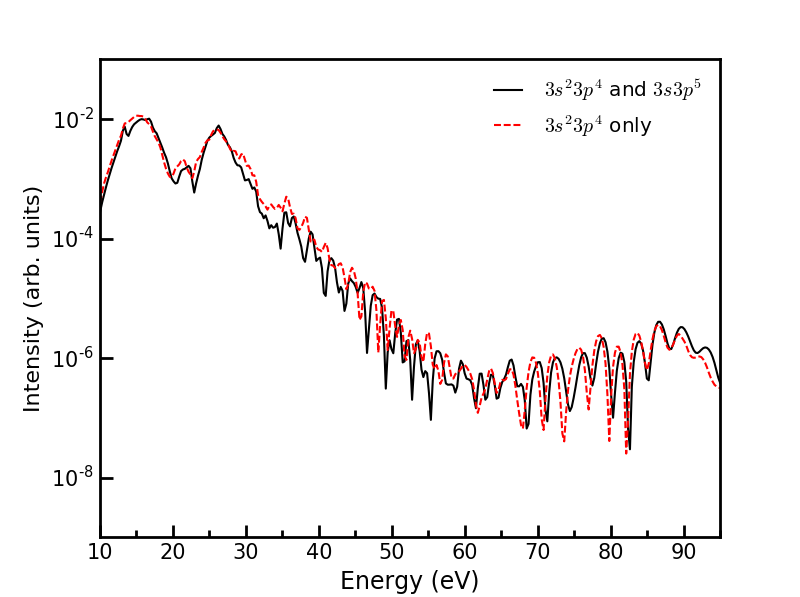}
\caption{(Color online) Harmonic spectra generated by the Ar$^{+}$ ion, subject to a two-color irradiation scheme.\ This comprises an $800\textrm{ nm}$, $3\times 10^{14}\,\textrm{Wcm}^{-2}$, IR laser pulse, in combination with a $92\textrm{ nm}$, $3\times 10^{12}\,\textrm{Wcm}^{-2}$, XUV pulse.\ The latter is timed to coincide with the penultimate maximum of the IR field.\ The yields are compared for two ionic structure configurations, in which all Ar$^{2+}$ $3s^{2}3p^{4}$ and $3s3p^{5}$ ionization thresholds are retained (solid, black line), or only the $3s^{2}3p^{4}$ thresholds (dashed, red line).}
\label{fig:thresholds}
\end{figure}

Figure \ref{fig:thresholds} compares the two-color spectra, in a restricted energy range, as generated with the aforementioned ionic structure configurations.\ We find that the gross features of the spectrum are largely preserved upon precluding ionization of a $3s$ electron.\ The latter confirms that both the plateau, and cutoff, harmonics arise primarily from the action of a $3p$ electron.\ Above the target ionization threshold, calculations neglecting the emission of a $3s$ electron display a moderately enhanced yield (typically less than an order of magnitude).\ Importantly, the energy range $35\textrm{ eV}$ to $55\textrm{ eV}$ is spanned by two Rydberg series, converging onto the $3s3p^{5}\, {^{3}P^{o}}$ and $3s3p^{5}\, {^{1}P^{o}}$ Ar$^{2+}$ ionization thresholds (see \cite{BrownvanderHart2012} for the precise threshold energies).\ We can thus assume that the reaction of multiple electrons, to the laser field, actually hinders HHG in the present scheme.\

Given the dominance of the ${^{1}D^{e}}$ and $^{1}S^{e}$ thresholds for the Ar$^{+}$ harmonic response (see Section \ref{subsec:competition}), we expect that the XIHHG excitation/tunnelling mechanism will be mediated by substantial populations in the $3s^{2}3p^{4}\, ({^{1}D^{e}})nl$ and $3s^{2}3p^{4}\, ({^{1}S^{e}})nl$ Rydberg states.\ These may couple to the $3s3p^{5}\, ({^{1}P^{o}})nl$ states via an IR- or XUV-driven, single-electron $3s\rightarrow3p$ transition.\ Rydberg series converging onto the $3s3p^{5}$ thresholds effectively describe the excitation of an inner-valence $3s$ electron, whilst the harmonic response is primarily determined by the action of an outer-valence $3p$ electron.\ The existence of such distinct pathways for ionic HHG, competing for electronic population, has a suppressive effect on the overall yield.\

As for the above-threshold harmonics, we find only minor deviations in yield (less than a factor of 3 on average) below the ionization threshold.\ In particular, the dipolar response at energies comparable to the $3s3p^{6} \, {^{2}S^{e}}$ excitation threshold (around $13.5\textrm{ eV}$) appears largely independent of the presence, or absence, of the $3s3p^{5}nl$ Rydberg series.\ This echoes the conclusions of the preceding section, wherein it had been established that the ${^{2}S^{e}}$ state plays no essential role in the XIHHG process.\ These findings suggest that, despite the choice of XUV pulse parameters (especially the central frequency), we cannot selectively resolve the dynamics of the $3s$ electron under these conditions.\ The HHG process is dictated by the ionization and recollision dynamics of a $3p$ electron, with the XUV field merely promoting the former through provision of high-energy photons.\ Inner-valence emission channels are manifest only in their suppressive effect for the total harmonic yield, which is ultimately mediated by interference amongst competing excitation pathways.\

\subsection{Time delay Scan of the Harmonic Response}

Having confirmed that a combination of XUV and IR laser light, with appropriately chosen time delay, can afford enhanced HHG yields from Ar$^{+}$, we now consider the effect of varying the time delay between the pulses.\ In doing so, we may assess the implications, for the HHG yield, of a more systematic control over the timing of the initial ionization event.\

\begin{figure}[h!]
\subfloat[]{%
  \includegraphics[height=8cm,width=1.0\columnwidth]{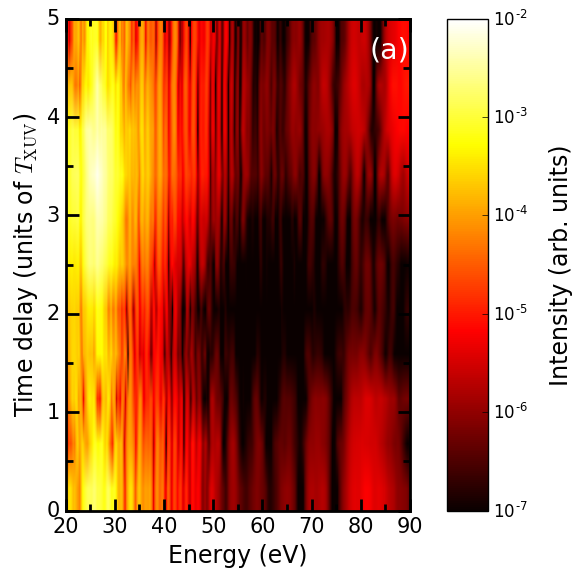}%
}\\
\subfloat[]{%
  \includegraphics[height=7cm,width=\columnwidth]{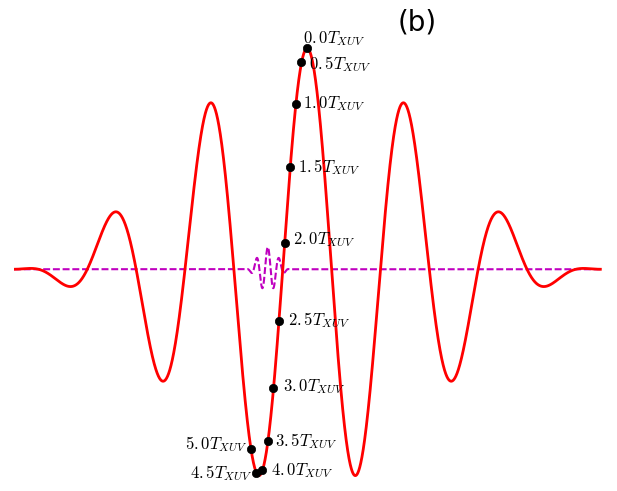}%
}
\caption{(Color online) Time delay scan of the harmonic response of Ar$^{+}$, subject to a combination of IR ($800\textrm{ nm}$, $3\times 10^{14}\,\textrm{Wcm}^{-2}$) and XUV ($92\textrm{ nm}$, $3\times 10^{12}\,\textrm{Wcm}^{-2}$) laser light.\ (a) Evolution of the two-color harmonic spectrum with time delay.\ (b) Two-color field configuration, with separate depiction of the XUV (dashed, magenta line) and IR (solid, red line) pulses.\ Filled circles (black) indicate, for specified time delays, the IR field when the XUV pulse attains peak intensity.\ All time delays are measured between the peaks of the XUV and IR pulses, and expressed in units of the XUV pulse period, $T_{\textrm{XUV}}\approx 0.31\textrm{ fs}$.}
\label{fig:comp}
\end{figure}

Figure \ref{fig:comp} presents a time delay scan of the Ar$^{+}$ harmonic response.\ Perhaps the most notable effect is the enhancement in the harmonic yield both above, and somewhat below, the ionization threshold, for time delays satisfying $0.0T_{\textrm{XUV}} < \Delta < 1.0T_{\textrm{XUV}}$ and $3.0T_{\textrm{XUV}} < \Delta < 4.0T_{\textrm{XUV}}$.\ In these cases, the central peak of the XUV pulse is near-coincident in time with either the primary, or penultimate, peak of the IR field.\ The XIHHG mechanism, discussed previously in the case $\Delta = 3.5T_{\textrm{XUV}}$, is then most efficient.\ Multiphoton absorption, from the XUV field, can drive direct photoionization from the $3p$ subshell, or excite the ion into high-lying $3s^{2}3p^{4}nl$ Rydberg states, from which the intense IR field can mediate tunnel ionization.\ Multiphoton processes, enabled by one or both laser fields, thereby enhance dipolar emission throughout the plateau region, and especially around the $3p$ ionization threshold.\

The efficacy of the XIHHG excitation/tunnelling process, for time delays that fall within the aforementioned ranges, accounts for the pronounced intensities of the plateau harmonics.\ In comparison, for $1.0T_{\textrm{XUV}} < \Delta < 2.0T_{\textrm{XUV}}$, XUV-driven transitions are not followed with significant tunnel ionization, by virtue of the decreasing intensity of the IR field.\ We thus expect a reduction in the degree of ionization, and consequently, a reduction in the harmonic yield.\ Moreover, by ejecting an electron when the field is weak, we fail to promote energetic recollision trajectories, thereby compromising the efficiency of the HHG process.\

The timing of the XUV pulse is important in yet another respect.\ When the XUV pulse temporally overlaps the primary peak of the IR field, as for $0.0T_{\textrm{XUV}} < \Delta < 1.0T_{\textrm{XUV}}$, the cutoff appears around $81\textrm{ eV}$.\ However, for somewhat longer time delays, spanning $3.0T_{\textrm{XUV}} < \Delta < 4.0T_{\textrm{XUV}}$, the cutoff emerges around $90\textrm{ eV}$.\ We claim that this is a consequence of strong-field dynamics alone, and provide confirmation by simulating the classical, laser-driven motion of an ejected electron.\

\begin{figure}
\includegraphics[width=0.50\textwidth]{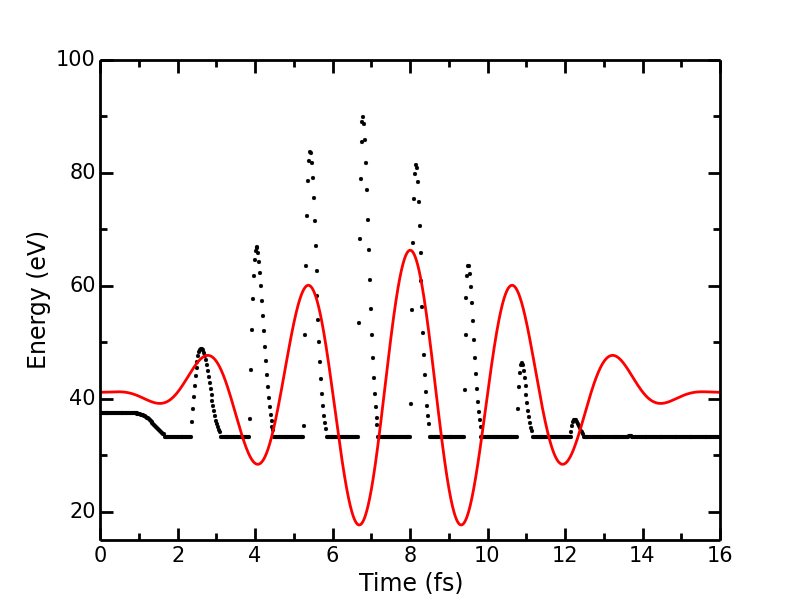}
\caption{(Color online) Harmonic photon energies (black, filled circles), arising from ionization at given instants of time, in an $800\textrm{ nm}$, $3\times 10^{14}\,\textrm{Wcm}^{-2}$, IR laser pulse (solid, red line), assuming a classical trajectory model.}
\label{fig:harm_energ}
\end{figure}

Figure \ref{fig:harm_energ} presents the distribution of harmonic photon energies, assuming ionization at different instants of time in the IR field (the weak XUV field modifies the recollision energies negligibly).\ If the XUV pulse excites the target into a (superposition of) Rydberg state(s), its susceptibility to tunnel ionization is increased.\ Since, for $0.0T_{\textrm{XUV}} < \Delta < 1.0T_{\textrm{XUV}}$, this occurs around the primary peak of the IR pulse, we might anticipate that the ionization response there dominates.\ From Figure \ref{fig:harm_energ}, we find that an electron released into the field near this peak will recollide with the parent ion, but give rise to a photon energy no higher than about $81\textrm{ eV}$.\ Thus, short IR time delays promote only sub-optimal recollision trajectories, leading to lower cutoff energies, and hence shorter plateau structures.\ In contrast, for $3.0T_{\textrm{XUV}} < \Delta < 4.0T_{\textrm{XUV}}$, ionization is enhanced within an optimal, sub-cycle time interval (about the penultimate IR peak), giving rise to the most energetic recollisions possible under the prevalent field conditions.\ This yields higher cutoff energies, and longer harmonic plateaus.\ Note that, in both cases, the XUV-mediated ionization affords enhanced intensities in their respective cutoff regions.\ Otherwise, the timing of the XUV pulse limits the efficiency of the XIHHG process, as reflected by lower cutoff yields.

There is some current interest in the characterization of sub-threshold harmonics, given that the details of the target atomic structure can be imprinted on the spectral emission \cite{BrownvanderHart2012,Brownetal2012,Soiferetal2010,Xiongetal2017}.\ The use of combined (XUV + IR) fields has traditionally been the concern of multi-dimensional absorption spectroscopies \cite{Dingetal2016,Wangetal2010,Holleretal2011,Bernhardtetal2014}, but Figure \ref{fig:comp} suggests that such a scheme may also effect non-negligible changes in the harmonic response, especially below the target ionization threshold.\ We observe that, for time delays satisfying $\Delta \approx 0.0T_{\textrm{XUV}}$ and $3.0T_{\textrm{XUV}} < \Delta < 4.0T_{\textrm{XUV}}$, the harmonics which span the range $20\textrm{ eV}$ to $25\textrm{ eV}$ accumulate considerable intensity.\ This is an unexpected feature of the present results.\ The three-step (recollision) process should not contribute to the yield of sub-threshold harmonics, being initiated only at energies exceeding the ionization threshold.\ The characteristics of low-order harmonics are often determined by dipole-allowed transitions amongst resonantly-coupled bound states.\ We propose that low-lying $3s^{2}3p^{4}nl$ Rydberg states, spanning the aforementioned energy range, become resonantly populated through XUV, or (XUV + IR), few-photon absorption.\ When the XUV pulse is coincident with a peak of the IR field, the latter may stimulate the ion to de-excite, with radiative emission.\ When the timing of the XUV pulse is otherwise, these Rydberg states do not depopulate so efficiently, and the lower-order harmonics exhibit reduced intensities.\

\section{Conclusions}

We have applied {\it ab initio} RMT theory to investigate XIHHG in Ar$^{+}$.\ With an appropriate choice of time delay between the XUV and IR pulses, we find that XUV-mediated multiphoton processes afford substantial enhancements in the yield of plateau harmonics, including those below the target ionization threshold.\ The single-color spectrum presents an important signature of the target electronic structure, in the form of a pronounced Cooper minimum.\ The latter has previously been observed in photoionization spectra of Ar$^{+}$, reinforcing the notion that HHG could enable a novel mode of spectroscopy.\ This minimum also persists in the two-color spectrum, suggesting that the XIHHG scheme we propose retains the spectroscopic potential of single-field HHG, whilst offering improved conversion efficiencies.\ In general, the measurement of ionic harmonic yields may facilitate the retrieval of their photoionization, and photorecombination, cross-sections throughout a broad range of energies, as demonstrated for neutral atoms \cite{Shineretal2011}.\

The competition amongst low-lying ionization thresholds, whose separation is comparable to the IR photon energy, has long been of interest in atomic HHG.\ We have shown that multi-color field configurations can alter the relative importance of different electron emission channels.\ Specifically, by driving HHG from an XUV-excited state, the effect of orbital alignment in the ground state, which ensures a weaker excitation/ionization response of $m = \pm 1$, as opposed to $m = 0$, electrons can be made less significant.\ In HHG from Ar$^{+}$, this leads to an enhanced population of channels connected with the ${^{3}P^{e}}$ threshold, accessible only by $m = \pm 1$ emission.\ The present findings could, however, be considerably more far-reaching for the control of ultrafast and strong-field processes.\ In molecules, for instance, excitation of particular electronic or vibrational states, via an appropriate multi-color irradiation strategy, could contribute to the enhancement of selected ionization, dissociation or even HHG pathways.\ This would aid in the manipulation of complex, multichannel dynamics in such systems.\

We have examined the potential role of an intermediate resonance for the present XIHHG scheme.\ Excitation of the low-lying $3s3p^{6} \, {^{2}S^{e}}$ state was effected via an XUV-driven, $3s\rightarrow 3p$ transition.\ However, little population transfer was achieved, and no signatures of HHG from this state were observed.\ Higher XUV peak intensities may enable a more effective participation of such a state in the HHG process.\ Alternatively, schemes surrounding the excitation of outer-valence electrons, and high-lying Rydberg states, could prove more fruitful \cite{Beaulieuetal2016}.\

Additionally, the general consequences of multielectron interference effects, involving the outer-shell $3s$ and $3p$ electrons, have been highlighted.\ Selective closure of $3s$ emission channels incurs only a minor enhancement in the yield of plateau harmonics, indicating that competing excitation pathways actually hinder HHG in the present scheme.\ Our findings seed doubt that such weak effects could be detected in an experimental realization.\  The relatively low probability of ionizing an inner-valence electron ensures that, even with a tuned XUV pulse, the outer-valence electrons dominate the HHG process.\ To resolve inner-valence excitations, we must devise more carefully a multi-color irradiation scheme, with appropriately selected spectral and temporal properties \cite{BrownvanderHart2016}.\

The importance of a systematic control, over the timing of the initial ionization event, has been demonstrated for XIHHG in Ar$^{+}$.\ In particular, we have confirmed that XUV multiphoton processes confer the greatest increases in yield whenever the XUV pulse coincides, in time, with a peak of the IR pulse.\ Moreover, depending on the phase of the IR field when ionization is promoted, the harmonic cutoff energy may be varied.\ Interestingly, the characteristics of the harmonics below the target ionization threshold, which are often dictated by intricate, bound state dynamics, also appear somewhat sensitive to the time delay between the pulses.\ These findings suggest that (XUV + IR) irradiation schemes, traditionally employed in absorption spectroscopy techniques, may also hold considerable promise for multi-dimensional harmonic spectroscopy.\

An experimental implementation of the XIHHG scheme proposed here presents several challenges.\ Not least among these is the preparation of the Ar$^{+}$ target.\ However, we note that in past experimental works, the highest-order harmonics from laser-irradiated Ar have indeed been attributed to Ar$^{+}$ \cite{Gibsonetal2004,Zepfetal2007}.\ These arise from a sequential process, in which Ar first ionizes, and then undergoes HHG.\ Moreover, although ionization in a linearly polarized field will leave Ar$^{+}$ predominantly in the $M=0$ state, the ion will subsequently evolve under the spin-orbit interaction.\ As a consequence, after a time of the order of $10\textrm{ fs}$, a significant $M=1$ population will arise.\ Previous R-matrix studies have demonstrated that enhanced HHG can be achieved with $M=1$ alignment, in both Ne$^{+}$ \cite{Hassounehetal2014} and Ar$^{+}$ \cite{BrownvanderHart2013}.\ It would thus be of considerable interest to explore the role of spin-orbit dynamics in ionic XIHHG schemes, especially those in which variable XUV pulse time delays play a fundamental role.\ Efforts to incorporate spin-orbit coupling, within the RMT formalism, are on the horizon.\

Furthermore, the laser regime which optimizes the resolution of XUV-initiated dynamics is not yet well understood.\ Our previous work has shown that long-wavelength driving pulses hold much promise, reducing the direct contribution of the IR field to the harmonic spectrum \cite{BrownvanderHart2016}.\ Although this was formerly thought untenable -  most experiments in attosecond physics employ the ubiquitous ($800\textrm{ nm}$) Ti:sapphire laser - a recent report has demonstrated the stable production of ultrashort, XUV pulses with a $1.8\,\mu\textrm{m}$ driver \cite{Gaumnitzetal2017}.\ Such developments enhance the possibility that spectroscopic XIHHG strategies, similar to the one outlined in this article, might become experimentally viable within the next several years.\

\begin{acknowledgements}
DDAC acknowledges financial support from the UK Engineering and Physical Sciences Research Council (EPSRC).\ HWvdH and ACB also acknowledge financial support from EPSRC, under Grant No.\ EP/P013953/1.\ This work relied on the ARCHER UK National Supercomputing Service (\href{http://www.archer.ac.uk/}{\texttt{www.archer.ac.uk}}).\ The data presented in this paper may be found using Ref.\ \cite{PUREdatasets}.\

\end{acknowledgements}

\end{document}